\documentstyle[12pt,epsfig,twoside,rotating]{article}
\epsfverbosetrue
\newcommand{\beq}{\begin{equation}}
\newcommand{\eeq}{\end{equation}}
\newcommand{\beqn}{\begin{eqnarray}}
\newcommand{\eeqn}{\end{eqnarray}}
\newcommand{\beqns}{\begin{eqnarray*}}
\newcommand{\eeqns}{\end{eqnarray*}}
\newcommand{\vs}{\\[0.3cm]\indent}

\newcommand{\hm}{\hspace{-0.05cm}}

\newcommand{\intl}{\int\limits}
\newcommand{\ointl}{\oint\limits}

\newcommand{\e}{\epsilon}
\def\NP{{\it Nucl. Phys.}}
\def\PL{{\it Phys. Lett.}}
\def\PR{{\it Phys. Rev.}}
\def\PRep{{\it Phys. Rep.}}
\def\PRL{{\it Phys. Rev. Lett.}}

\def\ZP{{\it Z. Phys.}}
\def\EPJ{{\it Europ. Phys. J.}}

\def\ea{{\it et al.}}
\def\Cl{Collaboration}
\def\MeVE{~MeV}
\def\pc{$\%$}

\def\sf{spectral function}

\def\ee{$e^+e^-$}

\def\aqed{$\alpha(s)$}
\def\aqedZ{$\alpha(M_{\rm Z}^2)$}

\def\daqedhZ{$\Delta\alpha_{\rm had}(M_{\rm Z}^2)$}
\def\amuhad{$a_\mu^{\rm had}$}
\def\aehad{$a_e^{\rm had}$}
\def\MSbar{$\overline{\rm MS}$}
\def\FOPTCI{$\rm FOPT_{\rm CI}$}
\def\ie{{\it i.e.}} 
\def\eg{{\it e.g.}} 
\def\via{via} 

\def\rs{\raisebox{1.5ex}[-1.5ex]}

\begin{document}

\begin{titlepage}
\setcounter{page}{1}

\begin{flushright} 
{\bf LAL 98-38}\\
May 1998
\end{flushright} 

\begin{center}
\vspace{1cm}
{\Large
  {\LARGE N}EW {\LARGE R}ESULTS ON THE {\LARGE H}ADRONIC 
  {\LARGE C}ONTRIBUTIONS \\[0.3cm]
    TO $\alpha(M_{\rm Z}^2$) AND TO $(g-2)_{\mu}$ \\
}
\vspace{1.5cm}
\begin{large}
Michel Davier\footnote{E-mail: davier@lal.in2p3.fr}
and Andreas H\"ocker\footnote{E-mail: hoecker@lal.in2p3.fr} \\
\end{large}
\vspace{0.5cm}
{\small \em Laboratoire de l'Acc\'el\'erateur Lin\'eaire,\\
IN2P3-CNRS et Universit\'e de Paris-Sud, F-91405 Orsay, France}\\
\vspace{3.5cm}

{\small{\bf Abstract}}
\end{center}
{\small
\vspace{-0.2cm}
We reevaluate the dispersion integrals of the leading order hadronic 
contributions to the running of the QED f\/ine structure constant \aqed\ at 
$s=M_{\rm Z}^2$, and to the anomalous magnetic moments of the muon and the
electron. Finite-energy QCD sum rule techniques complete the data from \ee\ 
annihilation and $\tau$ decays at low energy and at the $c\bar{c}$ 
threshold. Global quark-hadron duality is assumed in order to 
resolve the integrals using the Operator Product Expansion wherever
it is applicable. We obtain \daqedhZ\,$=(276.3\pm1.6)\times10^{-4}$ 
yielding $\alpha^{-1}(M_{\rm Z}^2)=128.933\pm0.021$, and
$a_\mu^{\rm had}=(692.4\pm6.2)\times 10^{-10}$ with which we
f\/ind for the complete Standard Model prediction
$a_\mu^{\rm SM}=(11\,659\,159.6\pm6.7)\times10^{-10}$.
For the electron, the hadronic contribution reads
$a_e^{\rm had}=(187.5\pm1.8)\times 10^{-14}$.
The following formulae express our results on the running of $\alpha$ 
at $M_{\rm Z}^2$ as a function of the input value for $\alpha_s(M_{\rm Z}^2)$ 
and its error:
\beqns
    \Delta\alpha_{\rm had}(M_{\rm Z}^2) 
        &=& 249.8 + 221\alpha_s(M_{\rm Z}^2) 
            \pm 1.5 \pm 221 \Delta\alpha_s(M_{\rm Z}^2)~, \\
    \alpha^{-1}(M_{\rm Z}^2)
        &=& 129.297 - 3.03\alpha_s(M_{\rm Z}^2)
            \pm 0.020 \pm 3 \Delta\alpha_s(M_{\rm Z}^2)~.
\eeqns
\noindent
}
\vspace{2cm}
\vfill
\centerline{\it (Submitted to Physics Letters B)}
\vspace{1cm}
\thispagestyle{empty}

\end{titlepage}

\newpage\thispagestyle{empty}{\tiny.}\newpage
\setcounter{page}{1}

%
%
\section*{\it Introduction }

The running of the QED f\/ine structure constant $\alpha(s)$
and the anomalous magnetic moment of the muon are observables whose 
theoretical precisions are limited by second order loop ef\/fects
from hadronic vacuum polarization. Both magnitudes are related \via\
dispersion relations to the hadronic production rate $R$ in \ee\
annihilation. While far from quark thresholds and at suf\/f\/iciently 
high center-of-mass energy $\sqrt{s}$, $R(s)$ can be predicted by 
perturbative QCD, theory may fail when resonances occur, \ie, local 
quark-hadron duality is broken. Fortunately, one can circumvent this 
drawback by using \ee\ annihilation data for $R(s)$ and, as proposed in 
Ref.~\cite{g_2pap}, hadronic $\tau$ decays benef\/itting from the 
conserved vector current (CVC). With help from a moment analysis
using weighted integrals over low-energy \ee\ cross sections we showed
in Ref.~\cite{alphapap} that the Operator Product Expansion 
(OPE)~\cite{wilson} (also called SVZ approach~{\cite{svz}) can safely
be applied down to energies of $\sqrt{s}=1.8~{\rm GeV}$. It turned 
out that at this energy nonperturbative contributions to the dispersion 
integrals are very small.
\vs
In this letter, we improve our previous determinations by using f\/inite-energy
QCD sum rule techniques in order to access theoretically energy regions 
where perturbative QCD fails. This idea was recently advocated in 
Ref.~\cite{schilcher}. In principle, the method uses no additional 
assumptions beyond those applied in Ref.~\cite{alphapap}. However, 
parts of the dispersion integrals evaluated at low-energy and the
$c\bar{c}$ threshold are obtained from values of the Adler
$D$-function itself, where we assume local quark-hadron duality to hold.
We therefore perform an evaluation at rather high energies (3~GeV for 
$u,d,s$ quarks and 15~GeV for the $c$ quark contribution) to suppress
deviations from the local duality assumption by nonperturbative
phenomena. We present a thorough evaluation of the associated 
theoretical and experimental uncertainties.

%
%

\section*{\it Running of the QED Fine Structure Constant}

The running of the electromagnetic f\/ine structure constant \aqed\
is governed by the renormalized vacuum polarization function,
$\Pi_\gamma(s)$. For the spin 1 photon, $\Pi_\gamma(s)$ is given 
by the Fourier transform of the time-ordered product of the 
electromagnetic currents $j_{\rm em}^\mu(s)$ in the vacuum 
$(q^\mu q^\nu-q^2g^{\mu\nu})\,\Pi_\gamma(q^2)=
i\int d^4x\,e^{iqx}\langle 0|T(j_{\rm em}^\mu(x)j_{\rm em}^\nu(0))|0\rangle$.  
With $\Delta\alpha(s)=-4\pi\alpha\,{\rm Re}
\left[\Pi_\gamma(s)-\Pi_\gamma(0)\right]$ and 
$\Delta\alpha(s)=\Delta\alpha_{\rm lep}(s)+\Delta\alpha_{\rm had}(s)$,
which subdivides the running contributions into a leptonic and a 
hadronic part, one has
\beq
    \alpha(s) \:=\: \frac{\alpha(0)}
                         {1 - \Delta\alpha_{\rm lep}(s)
                            - \Delta\alpha_{\rm had}(s)}~,
\eeq
where $4\pi\alpha(0)$ is the square of the electron charge in the 
long-wavelength Thomson limit. 
\vs
For the case of interest, $s=M_{\rm Z}^2$, the leptonic contribution 
at three-loop order has recently been calculated to be~\cite{steinh}
\beq
   \Delta\alpha_{\rm lep}(M_{\rm Z}^2) = 314.97686\times10^{-4}~.
\eeq
Using analyticity and unitarity, the dispersion integral for the 
contribution from the hadronic vacuum polarization 
reads~\cite{cabbibo}
\beq\label{eq_int_alpha}
    \Delta\alpha_{\rm had}(M_{\rm Z}^2) \:=\:
        -\frac{\alpha(0) M_{\rm Z}^2}{3\pi}\,
         {\rm Re}\!\!\!\intl_{4m_\pi^2}^{\infty}\!\!ds\,
            \frac{R(s)}{s(s-M_{\rm Z}^2)-i\epsilon}~,
\eeq
and, employing the identity $1/(x^\prime-x-i\epsilon)_{\epsilon\rightarrow0}
={\rm P}\{1/(x^\prime-x)\}+i\pi\delta(x^\prime-x)$, the above
integral is evaluated using the principle value integration 
technique.

%
%
\section*{\it Muon Magnetic Anomaly}

It is convenient to separate the Standard Model prediction for the
anomalous magnetic moment of the muon,
$a_\mu^{\mathrm SM}\equiv(g-2)_\mu/2$, into its dif\/ferent contributions,
\beq
    a_\mu^{\mathrm SM} \:=\: a_\mu^{\mathrm QED} + a_\mu^{\mathrm had} +
                             a_\mu^{\mathrm weak}~,
\eeq
where $a_\mu^{\mathrm QED}=(11\,658\,470.6\,\pm\,0.2)\times10^{-10}$ is 
the pure electromagnetic contribution (see~\cite{krause1} and references 
therein), \amuhad\ is the contribution from hadronic vacuum polarization,
and $a_\mu^{\mathrm weak}=(15.1\,\pm\,0.4)\times10^{-10}
$~\cite{krause1,peris,weinberg} accounts for corrections due to
exchange of the weak interacting bosons up to two loops.
\vs
Equivalently to \daqedhZ, by virtue of the analyticity of the 
vacuum polarization correlator, the contribution of the hadronic 
vacuum polarization to $a_\mu$ can be calculated \via\ the dispersion 
integral~\cite{rafael}
\beq\label{eq_int_amu}
    a_\mu^{\mathrm had} \:=\: 
           \frac{\alpha^2(0)}{3\pi^2}
           \intl_{4m_\pi^2}^\infty\!\!ds\,\frac{K(s)}{s}R(s)~,
\eeq
where $K(s)$ denotes the QED kernel~\cite{rafael2}~,
\beq
      K(s) \:=\: x^2\left(1-\frac{x^2}{2}\right) \,+\,
                 (1+x)^2\left(1+\frac{1}{x^2}\right)
                      \left({\mathrm ln}(1+x)-x+\frac{x^2}{2}\right) \,+\,
                 \frac{(1+x)}{(1-x)}x^2\,{\mathrm ln}x~,
\eeq
with $x=(1-\beta_\mu)/(1+\beta_\mu)$ and $\beta_\mu=(1-4m_\mu^2/s)^{1/2}$.
The function $K(s)$ decreases monotonically with increasing $s$. It gives
a strong weight to the low energy part of the integral~(\ref{eq_int_amu}).
About 92\pc\ of the total contribution to \amuhad\ is accumulated at c.m. 
energies $\sqrt{s}$ below 1.8~GeV and 72\pc\ of \amuhad\ is covered by 
the two-pion f\/inal state which is dominated by the $\rho(770)$ 
resonance. Data from vector hadronic $\tau$ decays published by the 
ALEPH Collaboration~\cite{aleph_vsf} provide a precise spectrum for
the two-pion f\/inal state as well as new input for the lesser known 
four-pion f\/inal states. This new information improves signif\/icantly
the precision of the \amuhad\ determination~\cite{g_2pap}.

%
%
\section*{\it Theoretical Prediction of $R(s)$}

Using a method based on weighted integrals over the low-energy
\ee\ cross sections, we showed in Ref.~\cite{alphapap} that 
perturbative QCD is safely applicable for the evaluation
of the dispersion integrals~(\ref{eq_int_alpha}) and 
(\ref{eq_int_amu}) at energies above 1.8~GeV, leaving out the 
$c\bar{c}$ threshold region where resonances occur. We stress
that this approach uses global quark-hadron duality only. 
Deviations from local duality are averaged to a negligible
contribution when calculating the integral. However, a 
systematic uncertainty is introduced through the cut
at explicitly 1.8~GeV, \ie, non-vanishing oscillations
could give rise to a bias after integration. 
In order to estimate the associated systematic error, we 
f\/itted dif\/ferent oscillating curves to the (rather imprecise) 
data around the cut region and obtained the following estimates
\beqn
  \Delta(\Delta \alpha_{\rm had}(M_{\rm Z}^2))
        &=& 0.15\times10^{-4}~,  \nonumber \\
  \Delta a_\mu^{\rm had} &=& 0.24\times10^{-10}~,
\eeqn
from the comparison of the integral over the oscillating simulated
data to perturbative QCD. These numbers are added as systematic
uncertainties to the corresponding low-energy integrals.
\vs
In asymptotic energy regions we use the formulae of Ref.~\cite{kuhn1}
which include mass corrections up to order $\alpha_s^2$ to evaluate 
the perturbative prediction of $R(s)$ entering into the
integrals~(\ref{eq_int_alpha}) and (\ref{eq_int_amu}).

%
%
\section*{\it Theoretical Approach Using Low-Energy Spectral Moments }

Following the suggestion made in Ref.~\cite{schilcher}, we can 
write the following identity:
\beq
\label{eq_f}
     F = \intl_{4m_\pi^2}^{s_0}\!\!ds\,R(s)\left[f(s) - p_{n,m}(s)\right] 
         + \frac{1}{2\pi i}\!\!\ointl_{|s|=s_0}\!\!\!\frac{ds}{s}
           \left[P_{n,m}(s_0) - P_{n,m}(s)\right] D_{uds}(s)~,
\eeq
with $P_{n,m}(s)=\int_0^sdt\,p_{n,m}(t)$ and 
$f(s)=\alpha(0)^2K(s)/(3\pi^2s)$ for 
$F\equiv a_{\mu,\,[2m_\pi,\;\sqrt{s_0}]}^{\rm had}$,
as well as $f(s)=\alpha(0)M_{\rm Z}^2/(3\pi s(s-M_{\rm Z}^2))$ for 
$F\equiv \Delta\alpha_{\rm had}(M_{\rm Z}^2)_{[2m_\pi,\;\sqrt{s_0}]}$.
The contour integral runs counter-clockwise around the circle from 
$s=s_0-i\e$ to $s=s_0+i\e$. The regular functions $p_{n,m}(s)$ approximate 
the kernel $f(s)$ in order to reduce the contribution of the f\/irst 
integral in Eq.~(\ref{eq_f}) which has a singularity at $s=0$ and 
is thus evaluated using experimental data. The second integral in 
Eq.~(\ref{eq_f}) can be evaluated theoretically applying QCD. 
To suppress unphysical subtractions we use the Adler $D$-function,
def\/ined as $D_{f_i}(s)=-12\pi^2 s \,d\Pi_{f_i}(s)/ds$ and 
$R_{f_i}(s)=12\pi{\rm Im}\Pi_{f_i}(s + i\e)$ for the set of quark
f\/lavours $f_i$, to calculate theoretically the second integral in
Eq.~(\ref{eq_f}), rather than the correlator $\Pi_{f_i}$. Using the 
OPE, the D-function for massive quarks is given by~\cite{3loop,kataev,bnp}
\beqn
\label{eq_d}
   D_{f_i}(-s) &=& N_C\sum_f Q_f^2
            \Bigg\{ 1 + d_0\frac{\alpha_s(s)}{\pi}
                     + d_1\left(\frac{\alpha_s(s)}{\pi}\right)^{\!\!2}
                     + \tilde{d}_2\left(\frac{\alpha_s(s)}{\pi}\right)^{\!\!3} 
                     \nonumber \\
         & & \hspace{2.2cm} 
           -\; \frac{m_f^2(s)}{s}
                \left( 6 + 28\,\frac{\alpha_s(s)}{\pi}
                        + (295.1-12.3\,n_f)
                          \left(\frac{\alpha_s(s)}{\pi}\right)^{\!\!2}
                 \right)  \nonumber \\
         & & \hspace{2.2cm} 
           +\; \frac{2\pi^2}{3}\left(1 - \frac{11}{18}\frac{\alpha_s(s)}{\pi}
                         \right)\frac{\left\langle\frac{\alpha_s}{\pi} 
                                           GG\right\rangle}{s^2} 
      \nonumber \\
         & & \hspace{2.2cm} 
           + \;8\pi^2\left(1 - \frac{\alpha_s(s)}{\pi}
                  \right)\frac{\langle m_f\bar{q_f}q_f\rangle}{s^2}
           +  \frac{32\pi^2}{27}\frac{\alpha_s(s)}{\pi}
              \sum_k\frac{\langle m_k\bar{q_k}q_k\rangle}{s^2}
      \nonumber \\
         & & \hspace{2.2cm} 
           + \;12\pi^2\frac{\langle{\cal O}_6\rangle}{s^3}
            \;+\; 16\pi^2\frac{\langle{\cal O}_8\rangle}{s^4}
      \Bigg\}~, 
\eeqn
where additional logarithms occur when $\mu^2\neq s$ and $\mu$ 
being the renormalization scale\footnote
{
   The negative energy-squared in $D(-s)$ of Eq.~(\ref{eq_d})
   is introduced when continuing the Adler function from the spacelike
   Euclidean space, where it is originally def\/ined, to the timelike
   Minkowski space by virtue of its analyticity property.
}.
The coef\/f\/icients
of the perturbative part read $d_0=1$, $d_1=1.9857 - 0.1153\,n_f$,
$\tilde{d}_2=d_2+\beta_0^2\pi^2/48$ with $\beta_0=11-2n_f/3$ and $n_f$ 
the number of involved quark f\/lavours, and 
$d_2=-6.6368-1.2001\,n_f-0.0052\,n_f^2-1.2395\,(\sum_f Q_f)^2/N_C\sum_f Q_f^2$.
The nonperturbative operators in Eq.~(\ref{eq_d}) are the gluon condensate, 
$\langle(\alpha_s/\pi) GG\rangle$, and the quark condensates,
$\langle m_f\bar{q_f}q_f\rangle$. The latter obey approximately the PCAC 
relations
\beqn
\label{eq_pcac}
     (m_u + m_d)\langle\bar{u}u + \bar{d}d\rangle
       & \simeq & - 2 f_\pi^2 m_\pi^2~, \nonumber \\
     m_s\langle\bar{s}s\rangle 
       & \simeq & - f_\pi^2 (m_K^2-m_\pi^2)~,
\eeqn
with the pion decay constant $f_\pi=(92.4\pm0.26)$\MeVE~\cite{pdg}.
In the chiral limit the relations $f_\pi=f_K$ and
$\langle\bar{u}u\rangle=\langle\bar{d}d\rangle=\langle\bar{s}s\rangle$
hold.
The complete dimension $D=6$ and $D=8$ operators are parametrized 
phenomenologically using the vacuum expectation values  
$\langle{\cal O}_6\rangle$ and $\langle{\cal O}_8\rangle$, respectively.
\vs
The functions $p_{n,m}(s)$ introduced in Eq.~(\ref{eq_f}) are 
chosen in order to reduce the {\it uncertainty} of the data integral.
This approximately coincides with a low residual value of the 
integral, \ie, a good approximation of the integration kernel $f(s)$
by the $p_{n,m}(s)$. We use
\beq
\label{eq_pol}
    p_{n,m}(s) \equiv \frac{1}{(s + s_0 + \e)^m}\sum_{i=1}^n c_i
                      \left(1 - \left(\frac{s}{s_0}\right)^{\!\!i}\right)~,
\eeq
where the form $(1-s/s_0)$ ensures a vanishing integrand at
the crossing of the positive real axis where the validity of the OPE 
is questioned. Polynomials of order $s^n$ involve leading order
nonperturbative contributions of dimension $D=2(n+1)$. We therefore 
restrict the analysis to $n<2$. Including additional powers $n\ge2$ 
would require to f\/it the expectation values of the corresponding 
dimension $D\ge6$ operators as they are not known theoretically and a
direct application of the results on $\tau$ data~\cite{aleph_asf} is 
dangerous since deviations from the vacuum saturation hypothesis used 
could be $s$-dependent. The new information is then spoiled since
the additional f\/itted parameters deteriorate the accuracy of
\daqedhZ\ and \amuhad. Figure~\ref{fig_polfit} shows the
\begin{figure}[t]
  \epsfxsize12.cm
  \centerline{\epsffile{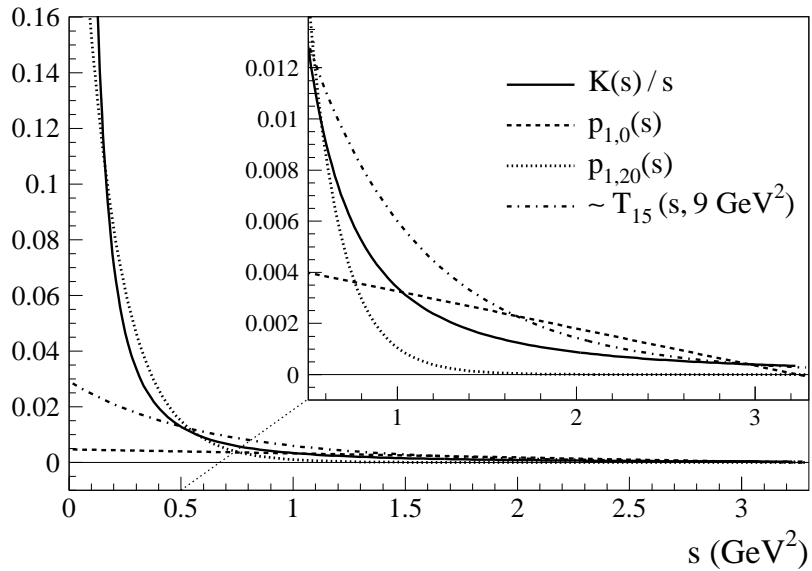}}
  \caption[.]{\label{fig_polfit}\it
              The integration kernel $K(s)/s$ of Eq.~\rm(\ref{eq_int_amu})
              \it and approximations according to the 
              definitions Eqs.~\rm(\ref{eq_pol}) \it and \rm(\ref{eq_momd}).}
\end{figure}
integration kernel of $a_\mu^{\rm had}$ and two of its $p_{n,m}$
approximations obtained from a $\chi^2$ minimization. At low energies
$p_{1,20}$ shows a clearly better agreement with the kernel than $p_{1,0}$
so that the uncertainty from the data integral in Eq.~(\ref{eq_f}) 
is reduced. However, setting $m>0$ increases dramatically the theoretical
uncertainties from $\alpha_s$, $m_s$ and the nonperturbative operators
which are then much larger than the errors of the pure data dispersion 
integrals. Therefore, we choose $m=0$ in the following analysis.
\vs
Several polynomials are used as spectral moments in order to adjust
$a_{\mu,\,[2m_\pi,\;\sqrt{s_0}]}^{\rm had}$ and 
$\Delta\alpha_{\rm had}(M_{\rm Z}^2)_{[2m_\pi,\;\sqrt{s_0}]}$ as well as the
gluon condensate simultaneously in a combined f\/it using 
Eq~(\ref{eq_f}). The experimental and theoretical correlations 
between the moments are calculated analytically from the total 
experimental covariance matrices and the variation of the
theoretical parameters within their uncertainties.

%
%
\section*{\it Theoretical Approach Using Dispersion Relations }
\label{sec_theodisp}

Another closely related method is the approximation of the 
integrals~(\ref{eq_int_alpha}) and (\ref{eq_int_amu})
via the dispersion relation of the Adler $D$-function
\beq
\label{eq_dispd}
    D_f(Q^2) = Q^2\!\!\intl_{4m_f^2}^\infty \!\!ds\,\frac{R_f(s)}{(s+Q^2)^2}~,
\eeq
for space-like $Q^2=-q^2$ and the quark f\/lavour $f$. Using the 
above integrand as approximation of the integration kernels in 
Eqs.~(\ref{eq_int_alpha}) and (\ref{eq_int_amu}), Eq.~(\ref{eq_f}) 
becomes
\beq
\label{eq_disp}
     F = \intl_{4m_\pi^2}^{s_0}\!\!ds\,R^{\rm Data}(s)
               \left[f(s) - \frac{A_FQ^2}{(s+Q^2)^2}\right] 
         + A_F\left(D_{uds}(Q^2) - Q^2\!\intl_{s_0}^\infty\!ds\,
           \frac{R_{uds}^{\rm QCD}(s)}{(s+Q^2)^2}\right)~,
\eeq
with a normalization constant $A_F$ to be optimized for both
$F\equiv a_{\mu,\,[2m_\pi,\;\sqrt{s_0}]}^{\rm had}$ and 
$F\equiv \Delta\alpha_{\rm had}(M_{\rm Z}^2)_{[2m_\pi,\;\sqrt{s_0}]}$.
The last integral in Eq~(\ref{eq_disp}) corrects the contribution
from the $D$-function and is calculated theoretically. One notices
that in contrast to Eq.~(\ref{eq_f}), where only global quark-hadron
duality is assumed, Eq.~(\ref{eq_disp}) requires local duality to hold
at $Q^2$. One therefore needs to choose $Q^2$ large against the scale
where nonperturbative phenomena appear. On the other hand, $Q^2$ too
large deteriorates the approximation of the kernel $f(s)$ in 
Eq.~(\ref{eq_disp}) enhancing the experimental uncertainty on $F$.
\vs
It is interesting to test whether derivatives of the Adler 
function~(\ref{eq_dispd}), def\/ined as
\beq
\label{eq_momd}
   M_n(Q^2) \equiv 
   \frac{(-1)^n}{(n+1)!} \left(Q^2\right)^{n+1}
   \frac{\partial^n\left(D_f(Q^2)/Q^2\right)}{\partial Q^{2n}}
   = \intl_{4m_f^2}^\infty\!\!ds\,R_f(s) T_n(s,Q^2)~,
\eeq
with $T_n(s,Q^2)=(Q^2)^{n+1}/(s+Q^2)^{n+2}$, ameliorate the evaluation
of the hadronic contributions~(\ref{eq_disp}). Higher derivatives $n>15$
indeed improve the approximation of the kernel function $f(s)$ by
$A_F T_n(Q^2)$ as can be seen from Fig.~\ref{fig_polfit}, and thus 
reduce the experimental uncertainty from the f\/irst integral 
on the r.h.s. of Eq.~(\ref{eq_disp}). Unfortunately, it
increases the theoretical errors of the $D$-function in 
Eq.~(\ref{eq_disp}). In order to demonstrate the ef\/fect we use 
the theoretical prediction~(\ref{eq_d}) which, setting
$\alpha_s(M_{\rm Z}^2)=0.1200$ and $\sqrt{Q^2}=3~{\rm GeV}$, yields
for $u,d,s$ quark f\/lavours the moments
\beqns
    M_0(9~{\rm GeV^2}) &=& 2.190 + 0.28\hat{m}_s^2
                          + 0.15\langle(\alpha_s/\pi) GG\rangle
                          + 1.8\langle m_q\bar{q}q\rangle
                          + 0.3\langle O_6\rangle
                          + 0.05\langle O_8\rangle~, \\
    M_4(9~{\rm GeV^2}) &=& 0.468 + 0.45\hat{m}_s
                          + 0.46\langle(\alpha_s/\pi) GG\rangle
                          + 5.3\langle m_q\bar{q}q\rangle
                          + 2.3\langle O_6\rangle
                          + 0.7\langle O_8\rangle~.
\eeqns
Taking as uncertainties:
$\Delta\alpha_s(M_{\rm Z}^2)=0.002$, 
$\Delta\hat{m}_s=0.07~{\rm GeV}$,
$\Delta\langle(\alpha_s/\pi) GG\rangle=0.02~{\rm GeV}^4$, 
$\Delta\langle m_q\bar{q}q\rangle=0.3\times10^{-4}~{\rm GeV}^4$,
$\Delta\langle O_6\rangle=0.01~{\rm GeV}^6$ and 
$\Delta\langle O_8\rangle=0.01~{\rm GeV}^8$, one obtains the relative
errors $\delta M_0(9~{\rm GeV^2})=0.8\%$ and
$\delta M_4(9~{\rm GeV^2})=7.9\%$, where the latter is far beyond the 
accuracy needed to improve the pure data results on \daqedhZ\ and 
\amuhad.

%
%
\section*{\it Theoretical Uncertainties}

Looking at Eq.~(\ref{eq_d}) it is instructive to subdivide the
discussion of theoretical uncertainties into three classes:
\begin{itemize}
  \item[(\it i)] {\it The perturbative prediction.} The estimation
                of theoretical errors of the perturbative series is
                strongly linked to its truncation at f\/inite order
                in $\alpha_s$. Due to the incomplete resummation of higher
                order terms, a non-vanishing dependence on the choice
                of the renormalization scheme (RS) and the 
                renormalization scale is left. Furthermore, one has
                to worry whether the missing four-loop order 
                contribution $d_3(\alpha_s/\pi)^4$ gives rise to
                large corrections to the perturbative series.
                On the other hand, these are problems to which 
                any measurement of the strong coupling constant 
                is confronted with, while their impact decreases with
                increasing energy scale. The error on the input parameter 
                $\alpha_s$ therefore ref\/lects the theoretical 
                uncertainty of the perturbative expansion in powers 
                of $\alpha_s$. A very robust $\alpha_s$ measurement
                is obtained from the global electroweak f\/it performed
                at the Z-boson mass where uncertainties from 
                perturbative QCD are reduced. The value found is 
                $\alpha_s(M_{\rm Z}^2)=0.120\pm0.003$~\cite{jerusalem}.
                A second precise $\alpha_s$ measurement is obtained 
                from the f\/it of the OPE to the hadronic width of the 
                $\tau$ lepton and to spectral moments~\cite{aleph_asf}, where
                the nonperturbative contribution was found to be lower than 
                $1\%$. Additional tests in which the mass scale was reduced 
                down to 1~GeV proved the excellent stability of the 
                $\alpha_s$ determination. The value recently 
                reported by the ALEPH Collaboration is
                $\alpha_s(M_{\rm Z}^2)=0.1202\pm0.0026$~\cite{aleph_asf}. 
                The consistency of the above values using quite dif\/ferent 
                approaches at various mass scales is remarkable and supports 
                QCD as the theory of strong interactions. Both measurements
                are almost uncorrelated so that we obtain the weighted
                average $\alpha_s(M_{\rm Z}^2)=0.1201\pm0.0020$, used in the
                following analysis.\par
                \hspace{0.4cm}
                Although contained in the above uncertainty
                of $\alpha_s$, we add the total dif\/ference between the 
                results obtained using contour-improved f\/ixed-order 
                perturbation theory (\FOPTCI) and FOPT (see comments in 
                Ref.~\cite{aleph_asf}) as systematic error.\par
                \hspace{0.4cm}
                Due to the truncation of the perturbative series, 
                the arbitrary choice of the renormalization scale $\mu$
                leaves an ambiguity. When setting $\mu^2\neq s$, additional
                logarithms enter the $D$-function~(\ref{eq_d}). We evaluate
                the corresponding uncertainty by varying $s\le\mu^2\le(3/2)s$.
  \item[(\it ii)] {\it The quark mass correction.} Since a theoretical 
                evaluation of the integrals~(\ref{eq_f}) and (\ref{eq_disp}) 
                is only applied far from quark thresholds, quark mass 
                corrections are small. We use the following settings:
                \beqns
                       m_{u,d}    &=& 0~,                  \\
                       m_s(1~{\rm GeV}) 
                                  &=& (0.20\pm0.07)~{\rm GeV}~,  \\
                       m_c(m_c)   &=& (1.3\pm0.2)~{\rm GeV}~,   \\
                       m_b(m_b)   &=& (4.2\pm0.2)~{\rm GeV}~,   \\
                       m_t(m_t)   &=& (176\pm6)~{\rm GeV}~.
                \eeqns
  \item[(\it iii)] {\it The nonperturbative contribution}. Using $n=1$ only,
                the functions~(\ref{eq_pol}) receive direct, 
                \ie, non-suppressed contributions from the dimension $D=4$ 
                terms. Associated nonperturbative parameters are the vacuum
                expectation values of the gluon and the quark condensates.
                While the latter can be obtained from the 
                PCAC relation~(\ref{eq_pcac}), for which a $20\%$ uncertainty
                is assumed, the gluon condensate cannot be f\/ixed
                theoretically. There fortunately exist experimental determinations
                using similar f\/inite-energy sum rule techniques which are almost 
                independent from the data used in this analysis: a f\/it
                using the $\tau$ vector plus axial-vector hadronic width and 
                spectral moments yields
   $\langle(\alpha_s/\pi)GG\rangle=(0.001\pm0.015)~{\rm GeV^4}$~\cite{aleph_asf}
                while a moment analysis using $c\bar{c}$ resonances results in 
   $\langle(\alpha_s/\pi)GG\rangle=(0.017\pm0.004)~{\rm GeV^4}$~\cite{reinders}.
                An adjustment on \ee\ data showed consistent 
                results~\cite{alphapap}. Following the above estimates, we use
                a gluon condensate of
                \beq
                   \langle(\alpha_s/\pi)GG\rangle=(0.015\pm0.020)~{\rm GeV^4}
                \eeq
                in this analysis.
\end{itemize}
Another tiny source of uncertainty is the error on the Z-boson mass.

%
%
\section*{\it Low-Energy Results}

Due to the suppression of nonperturbative contributions in 
powers of the energy scale $s$, the critical domain where 
nonperturbative ef\/fects may give residual contributions to
$R(s)$ is the low-energy regime with three active quark f\/lavours. 
However, we have shown in Ref.~\cite{alphapap} that at the scale 
$\sqrt{s_0}=1.8~{\rm GeV}$ global duality holds and the small 
nonperturbative ef\/fects are described by the OPE.
Up to this energy, $R(s)$ is obtained from the sum of the 
hadronic cross sections exclusively measured in the occurring 
f\/inal states. 
\vs
The data analysis follows exactly the line of Ref.~\cite{g_2pap}, 
In addition to the \ee\ annihilation data we use spectral functions 
from $\tau$ decays into two- and four f\/inal 
state pions measured by the ALEPH Collaboration~\cite{aleph_vsf}. 
As described in Ref.~\cite{g_2pap}, corrections to the charged
$\rho^\pm$ width have to be applied to account for small CVC-violating 
ef\/fects. The magnitude of the width dif\/ference,
$(\Gamma_{\rho^\pm}-\Gamma_{\rho^0})/\Gamma_{\rho}=(2.8 \pm 3.9)\times 10^{-3}$~,
translated into $a_\mu^{\rm had}$ and \daqedhZ\ is evaluated using a 
parametrization of the $\rho$ line shape based on vector 
resonances~\cite{aleph_vsf}. One thus obtains the additive corrections
\beqn
\label{eq_adif}
   \delta a_\mu^{\rm had} 
       &=& 
         -(1.3 \pm 2.0)\times10^{-10} \nonumber\\
   \delta \Delta\alpha_{\rm had}^{(5)}(M_{\rm Z}^2) 
       &=&
         -(0.09 \pm 0.12)\times10^{-4}
\eeqn
for the $\tau^-\!\rightarrow\pi^-\pi^0\nu_\tau$ \sf\ which is applied 
in the present (and former) analysis. Corrections for the higher 
mass resonances $\rho(1450), \rho(1700)$ are expected to be negligible. 
Extensive studies have been performed in Ref.~\cite{g_2pap} in order 
to bound unmeasured modes, such as ${\rm K\bar K}$ with pions
or the $\pi^+\pi^-4\pi^0$ f\/inal states, \via\ isospin constraints. 
We bring attention to the straightforward and statistically 
well-def\/ined averaging procedure and error propagation used 
in this paper as in the preceeding ones, which takes into account 
full systematic correlations between the cross section measurements. 
All technical details concerning the data analysis and the integration
method used are presented in Ref.~\cite{g_2pap}.
\vs
The experimental determination of the spectral moments in the f\/irst
integral of Eq.~(\ref{eq_f}) is performed as the sum over the respective 
moments of all exclusively measured \ee\ f\/inal states completed by 
$\tau$ data. 
\vs
We f\/it the following equations $(\sqrt{s_0}=1.8~{\rm GeV})$:
\beqn
\label{eq_fit_amu}
   \lefteqn{\frac{\alpha^2(0)}{3\pi^2}\hm\hm
            \intl_{4m_\pi^2}^{s_0}\!\!ds\,R^{\rm Data}(s)
      \left[\frac{K(s)}{s} - p_{1,0}^{(i)}(s)\right] \;=\;
       a_{\mu,\,[2m_\pi,\;\sqrt{s_0}]}^{\rm had}}\nonumber\\
   & &\hspace{2cm}
       \;-\; \frac{\alpha^2(0)}{6\pi^3 i}\!\!\ointl_{|s|=s_0}\!\!\frac{ds}{s}
                  \left[P_{1,0}^{(i)}(s_0) - P_{1,0}^{(i)}(s)\right] D_{uds}(s)~, 
        ~~~(i=1,\dots,9) \\[0.3cm]
\label{eq_fit_alpha}
  -\lefteqn{\frac{\alpha(0)M_{\rm Z}^2}{3\pi}\hm\hm
           \intl_{4m_\pi^2}^{s_0} \!\!ds\,R^{\rm Data}(s)
      \left[\frac{1}{s(s-M_{\rm Z}^2)} - p_{1,0}^{(i)}(s)\right] \;=\;
      \Delta\alpha_{\rm had}(M_{\rm Z}^2)_{[2m_\pi,\;\sqrt{s_0}]}}\nonumber\\
   & &\hspace{2cm}
           \;+\; \frac{\alpha(0)M_{\rm Z}^2}{6\pi^2 i}\!\!
             \ointl_{|s|=s_0}\!\!\frac{ds}{s}
                  \left[P_{1,0}^{(i)}(s_0) - P_{1,0}^{(i)}(s)\right] D_{uds}(s)~, 
        ~~(i=10,\dots,18) \\[0.3cm]
\eeqn    
where the polynomials $p_{1,0}^{(i)}$ are def\/ined in Eq.~(\ref{eq_pol}).
Table~\ref{tab_mom} shows the measured polynomial moments (l.h.s. of
Eqs.~(\ref{eq_fit_amu}) and (\ref{eq_fit_alpha})), together
with the f\/itted theoretical integrals (on the r.h.s. of
Eqs.~(\ref{eq_fit_amu}) and (\ref{eq_fit_alpha})) and the sum of both 
integrals for all 18 polynomials. 
\begin{table}[t]
\pagestyle{empty}
\setlength{\tabcolsep}{0.95pc}
{\small
\begin{tabular}{|cc|cc|cc|cc|c|} \hline 
Mom. & $c_1$ & Data & $\Delta$ & Theory & $\Delta$ 
  & Sum & $\Delta$ & $\sigma$\\
\hline\hline
(01)& 0        & 635.05 &  7.36 &    .00 &   .00 & 635.05 &  7.36 & .00 \\
(02)& .0005    & 601.49 &  6.93 &  33.47 &   .81 & 634.96 &  6.99 & .01 \\
(03)& .001     & 567.56 &  6.51 &  66.95 &  1.62 & 634.51 &  6.73 & .07 \\
(04)& .002     & 500.18 &  5.81 & 133.89 &  3.23 & 634.07 &  6.65 & .12 \\
(05)& .003     & 432.50 &  5.27 & 200.84 &  4.85 & 633.34 &  7.17 & .19 \\
(06)& .004     & 364.98 &  4.95 & 267.78 &  6.47 & 632.77 &  8.15 & .23 \\
(07)& .005     & 297.36 &  4.90 & 334.73 &  8.08 & 632.08 &  9.46 & .27 \\
(08)& .006     & 229.78 &  5.12 & 401.67 &  9.70 & 631.45 & 10.99 & .30 \\
(09)& .007     & 162.23 &  5.59 & 468.62 & 11.32 & 630.85 & 12.64 & .31 \\
\hline
(10)& 0        &  56.77 &  1.06 &    .00 &   .00 &  56.77 &  1.06 & .00 \\
(11)&$-.00005$ &  44.66 &   .86 &  11.98 &   .29 &  56.64 &   .90 & .12 \\
(12)&$-.00007$ &  39.80 &   .78 &  16.78 &   .41 &  56.58 &   .88 & .17 \\
(13)&$-.00009$ &  34.93 &   .70 &  21.57 &   .52 &  56.49 &   .87 & .24 \\
(14)&$-.00011$ &  30.08 &   .62 &  26.36 &   .64 &  56.44 &   .89 & .27 \\
(15)&$-.00013$ &  25.24 &   .55 &  31.15 &   .75 &  56.40 &   .93 & .28 \\
(16)&$-.00015$ &  20.46 &   .48 &  35.95 &   .87 &  56.40 &   .99 & .27 \\
(17)&$-.00017$ &  15.57 &   .41 &  40.74 &   .99 &  56.31 &  1.07 & .31 \\
(18)&$-.00019$ &  10.68 &   .35 &  45.53 &  1.10 &  56.21 &  1.16 & .35 \\
\hline
\end{tabular}
}
\caption[.]{\label{tab_mom}\it
            Polynomial moments: The horizontal lines separate the moments
            into Eqs.~\rm(\ref{eq_fit_amu}) \it and \rm(\ref{eq_fit_alpha})\it.
            The first column numbers the moments and the second defines the 
            coefficients corresponding to Eq.~\rm(\ref{eq_pol})\it. The 
            following columns give the l.h.s integrals of 
            Eqs.~\rm(\ref{eq_fit_amu}) \it and \rm(\ref{eq_fit_alpha})\it,
            which are the experimental results, and the r.h.s integrals,
            \ie, the theoretical values after fitting. The sums of both
            experimental and theoretical integrals give the values for
            $a_{\mu,\,[2m_\pi,\;1.8~{\rm GeV}]}^{\rm had}\times10^{10}$ and 
            $\Delta\alpha_{\rm had}(M_{\rm Z}^2)_{[2m_\pi,\;1.8~{\rm GeV}]}\times10^4$, 
            corresponding to the respective choice of polynomials.
            The last column shows the difference between the pure data 
            results (given in the first and the \rm10\it th line, 
            respectively, where the polynomial vanishes) and the sum
            in units of one standard deviation.}
\end{table}
\begin{table}[t]
\setlength{\tabcolsep}{0.15pc}
{\small
\begin{tabular}{cccccccccccccccccccccccccccc} \hline 
  Mom. &  (01) & (02) & (03) & (04) & (05) & (06) & (07) & (08) & (09) 
  & (10) & (11) & (12) & (13) & (14) & (15) & (16) & (17) & (18) 
\\ \hline
(1)
                                                            & 1 &   .97 &   .94
&   .82 &   .65 &   .46 &   .31 &   .19 &   .10 &   .75 &   .71 &   .66 &   .59
&   .51 &   .42 &   .34 &   .25 &   .18 &
 \\
(2) & --
                                                            & 1 &   .97 &   .88
&   .73 &   .57 &   .43 &   .31 &   .23 &   .71 &   .71 &   .68 &   .63 &   .57
&   .50 &   .42 &   .35 &   .29 &
 \\
(3) & -- & --
                                                            & 1 &   .93 &   .82
&   .68 &   .55 &   .44 &   .36 &   .66 &   .70 &   .69 &   .66 &   .62 &   .57
&   .51 &   .45 &   .39 &
 \\
(4) & -- & -- & --
                                                            & 1 &   .94 &   .86
&   .76 &   .68 &   .61 &   .50 &   .63 &   .66 &   .69 &   .69 &   .68 &   .65
&   .62 &   .59 &
 \\
(5) & -- & -- & -- & --
                                                            & 1 &   .95 &   .90
&   .85 &   .80 &   .31 &   .51 &   .59 &   .65 &   .70 &   .73 &   .74 &   .73
&   .73 &
 \\
(6) & -- & -- & -- & -- & --
                                                            & 1 &   .97 &   .94
&   .91 &   .13 &   .38 &   .49 &   .58 &   .66 &   .72 &   .76 &   .79 &   .80
&
 \\
(7) & -- & -- & -- & -- & -- & --
                                                            & 1 &   .97 &   .96
&   .00 &   .28 &   .40 &   .52 &   .62 &   .70 &   .76 &   .80 &   .82 &
 \\
(8) & -- & -- & -- & -- & -- & -- & --
                                                            & 1 &   .98 &  $-.10$
&   .19 &   .33 &   .45 &   .57 &   .67 &   .74 &   .79 &   .83 &
 \\
(9) & -- & -- & -- & -- & -- & -- & -- & --
                                                            & 1 &  $-.17$ &   .13
&   .27 &   .40 &   .53 &   .63 &   .72 &   .78 &   .82 &
 \\
(10) & -- & -- & -- & -- & -- & -- & -- & -- & --
                                                            & 1 &   .93 &   .86
&   .78 &   .67 &   .56 &   .44 &   .34 &   .24 &
 \\
(11) & -- & -- & -- & -- & -- & -- & -- & -- & -- & --
                                                            & 1 &   .97 &   .93
&   .87 &   .79 &   .70 &   .62 &   .54 &
 \\
(12) & -- & -- & -- & -- & -- & -- & -- & -- & -- & -- & --
                                                            & 1 &   .97 &   .93
&   .87 &   .80 &   .73 &   .66 &
 \\
(13) & -- & -- & -- & -- & -- & -- & -- & -- & -- & -- & -- & --
                                                            & 1 &   .97 &   .93
&   .88 &   .83 &   .77 &
 \\
(14) & -- & -- & -- & -- & -- & -- & -- & -- & -- & -- & -- & -- & --
                                                            & 1 &   .97 &   .94
&   .90 &   .86 &
 \\
(15) & -- & -- & -- & -- & -- & -- & -- & -- & -- & -- & -- & -- & -- & --
                                                            & 1 &   .97 &   .95
&   .92 &
 \\
(16) & -- & -- & -- & -- & -- & -- & -- & -- & -- & -- & -- & -- & -- & -- & --
                                                            & 1 &   .97 &   .96
&
 \\
(17) & -- & -- & -- & -- & -- & -- & -- & -- & -- & -- & -- & -- & -- & -- & --
& --                                                        & 1 &   .98 &
 \\
(18) & -- & -- & -- & -- & -- & -- & -- & -- & -- & -- & -- & -- & -- & -- & --
& -- & --                                                   & 1 &
 \\ \hline
\end{tabular}
}
\caption[.]{\label{tab_momcorr}\it
            Total experimental and theoretical correlations between the
            spectral moments used in the combined fit.}
\end{table}
The correlation matrix including experimental and theoretical 
correlations of the moments def\/ined in Table~\ref{tab_mom} is 
given in Table~\ref{tab_momcorr}. 
These correlations have been inserted into the $\chi^2$ minimization
used to f\/it the free varying parameters for which we obtain:
\beqn
\label{eq_fitres}
   a_{\mu,\,[2m_\pi,\;1.8~{\rm GeV}]}^{\rm had}
      &=& (634.3 \pm 5.6_{\rm exp} \pm 2.1_{\rm theo})\times 10^{-10}~, \nonumber\\
   \Delta\alpha_{\rm had}(M_{\rm Z}^2)_{[2m_\pi,\;1.8~{\rm GeV}]}
      &=& (56.53 \pm 0.73_{\rm exp} \pm 0.39_{\rm theo})\times 10^{-4}~,
\eeqn
yielding a minimum $\chi^2=1.0$ for 16 degrees of freedom.
The correlation between the f\/itted parameters~(\ref{eq_fitres}) 
amounts to $69\%$. 
\vs
The improvement in precision provided by this method can be easily
understood from Table~\ref{tab_mom}. As the chosen polynomial approaches
better the actual kernel, the uncertainty from the integral on data
is reduced, while the corresponding uncertainty from theory increases.
Some optimum is reached for some particular choice of polynomial,
providing the best constraint on the $a_\mu^{\rm had}$ and \daqedhZ\
determinations. It should be noted that the experimental precision
is not uniform throughout the energy range considered: for 
$\sqrt{s}\approx0.8~{\rm GeV}$, $R(s)$ is dominated by the $\rho$
spectral function, well measured in both $\tau$ decays and \ee\ 
annihilation, while at larger
$s$ values the uncertainties from some poorer measurements such as
$e^+e^-\rightarrow {\rm K}\bar{\rm K}\pi\pi$ or $6\pi$ take over.
We clearly disagree with the conclusions reached in Ref.~\cite{schilcher}
stating that the f\/inal uncertainty in the determination of 
\daqedhZ\ becomes insensitive to the experimental errors.
The experimental uncertainty can hardly be reduced to nothing even if 
the integral on data becomes very small. A vanishing error is only 
achieved if the polynomial approximates well the kernel, but
the theoretical uncertainty becomes prohibitive in this case.
This trend is illustrated in Table~\ref{tab_mom}.
\vs
As described before we may also use
Eq.~(\ref{eq_disp}) to constrain theoretically the low-energy
dispersion relations~(\ref{eq_int_alpha}) and (\ref{eq_int_amu}).
Here dif\/ferent moments are def\/ined by varying values for $Q^2$ 
and $A_F$. Unfortunately such moments are almost degenerate, \ie, they 
have very large correlations so that a combined f\/it as performed in
the case of the polynomials does not make much sense. We therefore
optimize the choice of $Q^2$ and $A_F$ in order to minimize the resulting
errors on $a_{\mu,\,[2m_\pi,\;\sqrt{s_0}]}^{\rm had}$ and
$\Delta\alpha_{\rm had}(M_{\rm Z}^2)_{[2m_\pi,\;\sqrt{s_0}]}$
setting again $\sqrt{s_0}=1.8~{\rm GeV}$. The corresponding results
are given in Table~\ref{tab_disp}. An analysis of the theoretical
uncertainties shows that at $\sqrt{Q^2}=3~{\rm GeV}$ nonperturbative
contributions to the $D$-function are negligible. The theoretical
errors given in Table~\ref{tab_disp} originate mainly from 
uncertainties of the massless perturbative series in Eq.~(\ref{eq_d}),
\ie, the error on $\alpha_s$ and the variation of $\mu$.
The weighted averages of pure data and the theory-improved results 
given take into account the correlations between both values.
These values agree with those from the combined f\/it of the 
polynomial moments. The most precise of the two evaluations are 
retained for the f\/inal results.

%
%
\section*{\it Results on the $c\bar{c}$ Threshold}

The $c\bar{c}$ threshold region involves the narrow resonances,
$J/\psi(1S)$, $\psi(2S)$ and $\psi(3770)$, parametrized using relativistic
Breit-Wigner formulae~\cite{eidelman,g_2pap}, as well as the broad 
resonance and 
\begin{table}[p]
\setlength{\tabcolsep}{0.6pc}
\begin{center}
{\normalsize
\begin{tabular}{lccc} \hline
   & $\Delta\alpha_{\rm had}(M_{\rm Z}^2)_{[2m_\pi,\;1.8~{\rm GeV}]}\times10^{4}$
      & $a_{\mu,\,[2m_\pi,\;1.8~{\rm GeV}]}^{\rm had}\times10^{10}$
\\ \hline\hline
$\sqrt{Q^2}$   & 3~GeV           & 3~GeV            \\
$A_F       $   & $-0.0003$ & 0.009                   \\
$I^{\rm exp}$ ($2m_\pi$ -- 1.8~GeV)
               & $44.67 \pm 0.70$ & $534.1 \pm 6.2$ \\
$D^{\rm theo}_{uds}(Q^2)$ 
               & $42.18 \pm 0.23$ & $353.5 \pm 2.0$ \\
$-I^{\rm theo}_{uds}$ (1.8~GeV -- $\infty$)
               & $-30.68 \pm 0.21$ & $-257.1 \pm 1.8$ \\
\hline
Total          & $56.17 \pm 0.70_{\rm exp} \pm 0.22_{\rm theo}$ 
                  & $630.5 \pm 6.2_{\rm exp} \pm 1.9_{\rm theo}$ \\
Data only      & $56.8 \pm 1.1^(\footnotemark[1]{^)}$ 
                                 & $635.1 \pm 7.4^(\footnotemark[1]{^)}$   \\
Correlation    & $95\%$          & $95\%$           \\
\hline
Average        & $56.36 \pm 0.70_{\rm exp}\pm0.18_{\rm theo}$ 
                 & $632.5 \pm 6.2_{\rm exp}\pm1.6_{\rm theo}$  \\
                    
\hline \\[0.5cm] \hline
   & $\Delta\alpha_{\rm had}
     (M_{\rm Z}^2)_{[\psi][3.7,\;5~{\rm GeV}]}\times10^{4}$
   & $a_{\mu,\,[\psi][3.7,\;5~{\rm GeV}]}^{\rm had}\times10^{10}$
\\ \hline\hline
$\sqrt{Q^2}$   & 15~GeV           & 15~GeV            \\
$A_F       $   & $-0.0009$ & 0.0015                  \\
$I^{\rm exp}$ (3.7 -- 5~GeV)
               & $7.02 \pm 0.45$ & $2.94 \pm 0.19$ \\
$I^{\rm exp}$ $(\psi(1S,2S,3770))$
               & $6.25 \pm 0.37$ & $6.12 \pm 0.35$  \\
$D^{\rm theo}_c(Q^2)$ 
               & $79.95 \pm 0.83$ & $37.22 \pm 0.38$ \\
$-I^{\rm theo}_c$ (5~GeV -- $\infty$)
               & $-74.01 \pm 0.31$ & $-34.55 \pm 0.15$ \\
$I^{\rm theo}_{uds}$ (3.7 -- 5~GeV)
               & $5.32 \pm 0.03$ & $2.47 \pm 0.02$ \\
\hline
Total          & $24.53 \pm 0.58_{\rm exp} \pm 0.88_{\rm theo}$ 
                   & $14.20 \pm 0.40_{\rm exp} \pm 0.40_{\rm theo}$ \\
                  
Data only (3.7 -- 5~GeV)      
               & $15.80 \pm 1.00^(\footnotemark[2]{^)}$ 
                                 & $6.93 \pm 0.44^(\footnotemark[2]{^)}$  \\
Data only $\psi(1S,2S,3770)$
               & $9.24 \pm 0.68^(\footnotemark[3]{^)}$ 
                                  & $7.51 \pm 0.44^(\footnotemark[3]{^)}$   \\
Correlation    & $55\%$          & $71\%$                  \\
\hline
Average        & $24.75 \pm 0.84_{\rm exp}\pm0.50_{\rm theo}$ 
                    & $14.31 \pm 0.50_{\rm exp} \pm 0.21_{\rm theo}$ \\
\hline 
\end{tabular}
{\footnotesize 
\parbox{14cm}
{
\vspace{0.2cm}
$^{1}\,$The slight modif\/ication compared to our previous values
       of $\Delta\alpha_{\rm had}(M_{\rm Z}^2)_{[2m_\pi,\;1.8~{\rm GeV}]}
       =56.9\times10^{-4}$ and $a_{\mu,\,[2m_\pi,\;1.8~{\rm GeV}]}^{\rm had}
       =636.5\times10^{-10}$~\cite{alphapap} are due to a reevaluation of 
       the contributions from $\tau$ data \\ \noindent
$^{2}\,$A reevaluation of the treatment of the experimental errors on 
       the contribution from charm threshold led to a signif\/icant 
       reduction of the very conservative uncertainties given in our 
       previous paper~\cite{alphapap} \\ \noindent
$^{3}\,$The changes in the resonance contributions compared to 
       Ref.~\cite{alphapap} is due to the multiplicative correction
       $(\alpha/\alpha(M_\psi^2))^2$, erroneously not applied 
       before~\cite{kuhnpriv}
}}
}
\end{center}
\caption[.]{\label{tab_disp}\it
            The optimized choices of the parameters $Q^2$ and $A_F$,
            the solutions of the corresponding integrals (``I'') in \rm
            Eq.~(\ref{eq_disp}) \it (upper table) and Eq.~\rm(\ref{eq_dispcc}) 
            \it (lower table) as well as the corresponding values for the
            Adler $D$-functions. Additionally given are the correlations
            between the theory-improved and the pure data results used
            to calculate the average of both.}
\vspace{0.5cm}
\end{table}
continuum region starting with the opening of the 
$e^+e^-\rightarrow D\bar{D}$ mode at about 3.7~GeV. A compilation of
the data used can be found in Ref.~\cite{g_2pap}. In this energy 
region we do not use the polynomial approach for which the precision
is limited by the uncertainty on the $c$-quark mass and the threshold
behaviour of the correlator. The f\/irst problem is less severe with the 
dispersion relation approach while the second one is inexistent. 
Equation~(\ref{eq_disp}) now reads:
\beqn
\label{eq_dispcc}
     F &=& \intl_{s_1}^{s_2} ds\,R(s)
               \left[f(s) - \frac{A_FQ^2}{(s+Q^2)^2}\right] 
         + \sum_{\psi=1S,2S,3770}\intl_0^\infty ds\,R_\psi(s)
               \left[f(s) - \frac{A_FQ^2}{(s+Q^2)^2}\right] \nonumber\\[0.3cm]
       & &  + A_F\left(D_c(Q^2) - Q^2\intl_{s_2}^\infty ds\,
              \frac{R_c^{\rm QCD}(s)}{(s+Q^2)^2}\right)
            + Q^2\intl_{s_1}^{s_2} ds\,
              \frac{R_{uds}^{\rm QCD}(s)}{(s+Q^2)^2}~,
\eeqn
where we choose for the continuum the integration ranges 
$\sqrt{s_1}=3.7~{\rm GeV}$ and $\sqrt{s_2}=5~{\rm GeV}$, and
set $F\equiv a_{\mu,\,[\psi][\sqrt{s_1},\;\sqrt{s_2}]}^{\rm had}$ and 
$F\equiv \Delta\alpha_{\rm had}(M_{\rm Z}^2)_{[\psi][\sqrt{s_1},\;\sqrt{s_2}]}$,
while $f(s)$ denotes the corresponding integration kernels,
respectively. Table~\ref{tab_disp} gives the results of the 
individual terms in Eq.~(\ref{eq_dispcc}) again after optimizing
the choice of $Q^2$ and the normalization $A_F$. The quoted error
on $D_c(Q^2)$ is dominated by the uncertainty on the $c$-quark mass.

%
%
\section*{\it Results for \daqedhZ\ and \amuhad}

In the previous sections we reevaluated the dispersion integrals
yielding the hadronic contributions to $a_\mu$ and to \aqedZ\
for the low-energy and $c\bar{c}$ threshold regions. According
to Ref.~\cite{alphapap} (see also Ref.~\cite{kuhnstein}),
we employ perturbative QCD using the formulae of Ref.~\cite{kuhn1}
for the cross section ratio $R(s)$ in other regions. For the
crossing of the $b\bar{b}$ threshold, we assume nonperturbative
ef\/fects to be negligible and use perturbative QCD in the context
of global duality (see, \eg, Ref.~\cite{pichjamin}) to evaluate the 
hadronic contributions. The theoretical error is then dominated by
the uncertainty on the $b$-quark production threshold $2M_b$, with
$M_b$ being the pole mass of the $b$-quark.
\vs
Table~\ref{tab_alphares} shows the experimental and theoretical 
evaluations of \daqedhZ, \amuhad\ and \aehad\ for the respective 
energy regimes\footnote
{
   The evaluation of \aehad\ follows the same procedure as \amuhad.
}. 
Experimental errors between dif\/ferent lines are
assumed to be uncorrelated, whereas theoretical errors but
those from $c\bar{c}$ and $b\bar{b}$ thresholds which are quark mass
dominated are added linearily.
\vs
According to Table~\ref{tab_alphares}, the combination of the theoretical
and experimental evaluations of the integrals~(\ref{eq_int_alpha}) 
and (\ref{eq_int_amu}) yields the f\/inal results
\beq
\label{eq_asres}
\begin{array}{|rcl|}
  \hline 
  & & \\
   ~~~\Delta\alpha_{\rm had}(M_{\rm Z}^2) 
      &=& (276.3 \pm 1.1_{\rm exp} \pm 1.1_{\rm theo})\times10^{-4}~~~ \\
  & & \\
   ~~~\alpha^{-1}(M_{\rm Z}^2) 
      &=& 128.933 \pm 0.015_{\rm exp} \pm 0.015_{\rm theo}~~~ \\ 
  & & \\
   ~~~a_\mu^{\rm had}
      &=& (692.4 \pm 5.6_{\rm exp} \pm 2.6_{\rm theo})\times10^{-10}~~~ \\
  & & \\
   ~~~a_\mu^{\rm SM}
      &=& (11\,659\,159.6 \pm 5.6_{\rm exp} \pm 3.7_{\rm theo})\times10^{-10}~~~ \\
  & & \\ 
  \hline
\end{array}
\eeq
and $a_e^{\rm had}=(187.5\pm1.7_{\rm exp}\pm0.7_{\rm theo})\times10^{-14}$ 
for the leading order hadronic contribution to $a_e$.
The total $a_\mu^{\rm SM}$ value includes an additional contribution
from non-leading order hadronic vacuum polarization summarized in 
Refs.~\cite{krause2,g_2pap} to be 
$a_\mu^{\rm had}[(\alpha/\pi)^3]=(-10.0\pm0.6)\times10^{-10}$.
\begin{table}[t]
\setlength{\tabcolsep}{0.2pc}
\begin{center}
{\small
\begin{tabular}{lccc} \hline \\[-0.33cm]
Energy~(GeV)
                         & $\Delta \alpha_{\rm had}(M_{\rm Z}^2)\times10^{4}$
                              & $a_\mu^{\rm had}\times10^{10}$ 
                                   & $a_e^{\rm had}\times10^{14}$ 
\\[0.07cm]
\hline \\[-0.33cm]
$(2m_\pi$ -- $1.8)_{uds}$
                         & $56.36\pm0.70_{\rm exp}\pm0.18_{\rm theo}$
                              & $634.3\pm5.6_{\rm exp}\pm2.1_{\rm theo}$ 
                                   & $173.67\pm1.7_{\rm exp}\pm0.6_{\rm theo}$ 
\\[0.07cm]
$(1.8$ -- $3.700)_{uds}$ & $24.53\pm0.28_{\rm theo}$  
                              & $33.87\pm0.46_{\rm theo}$     
                                   & $8.13\pm0.11_{\rm theo}$
\\[0.07cm]
$\psi(1S,2S,3770)_c$ && \\
$+~(3.7$ -- $5)_{udsc}$
                         &\rs{$24.75\pm0.84_{\rm exp}\pm0.50_{\rm theo}$}
                              & \rs{$14.31\pm0.50_{\rm exp}\pm0.21_{\rm theo}$} 
                                   & \rs{$3.41\pm0.12_{\rm exp}\pm0.05_{\rm theo}$} 
\\[0.07cm]
$(5$ -- $9.3)_{udsc}$    & $34.95\pm0.29_{\rm theo}$   
                              & $6.87\pm0.11_{\rm theo}$ 
                                   & $1.62\pm0.03_{\rm theo}$ 
\\[0.07cm]
$(9.3$ -- $12)_{udscb}$
                         &$15.70\pm0.28_{\rm theo}$
                              & $1.21\pm0.05_{\rm theo}$ 
                                   & $0.28\pm0.02_{\rm theo}$ 
\\[0.07cm]
$(12$ -- $\infty)_{udscb}$
                         & $120.68\pm0.25_{\rm theo}$   
                              & $1.80\pm0.01_{\rm theo}$ 
                                   & $0.42\pm0.01_{\rm theo}$ 
\\[0.07cm]
$(2m_t$ -- $\infty)_t$
                         &$-0.69\pm0.06_{\rm theo}$  
                              & $\approx0 $ 
                                   & $\approx0 $ 
\\[0.07cm]
\hline\\[-0.33cm]
$(2m_\pi$ -- $\infty)_{udscbt}$
                         & $276.3\pm1.1_{\rm exp}\pm1.1_{\rm theo}$
                              & $692.4\pm5.6_{\rm exp}\pm2.6_{\rm theo}$ 
                                   & $187.5\pm1.7_{\rm exp}\pm0.7_{\rm theo}$ 
\\[0.07cm]
\hline
\end{tabular}
}
\end{center}
\caption[.]{\label{tab_alphares}\it
            Contributions to \daqedhZ, \amuhad\ and to \aehad\ from the 
            different energy regions. The subscripts in the first column
            give the quark flavours involved in the calculation.}
\end{table}
Also the light-by-light scattering (LBLS) contribution has recently been 
reevaluated to be 
$a_\mu^{\rm had}[{\rm LBLS}]=(-7.9\pm1.5)\times10^{-10}$~\cite{kinolight}. 
Together with the value
$a_\mu^{\rm had}[{\rm LBLS}]=(-9.2\pm3.2)\times10^{-10}$~\cite{light2}, 
we use the average
$\langle a_\mu^{\rm had}[{\rm LBLS}]\rangle=(-8.5\pm2.5)\times10^{-10}$
so that the total higher order hadronic correction amounts to
$a_\mu^{\rm had}[(\alpha/\pi)^3+{\rm LBLS}]=(-18.5\pm2.6)\times10^{-10}$.
Figures~\ref{fig_results_alpha} and \ref{fig_results_amu} show a 
compilation of published results for the hadronic contributions 
to \aqedZ\ and $a_\mu$. Some authors give the hadronic contribution 
for the f\/ive light quarks only and add the top quark part separately. 
This has been corrected for in F\/ig.~\ref{fig_results_alpha}.

%
%
\section*{\it Conclusions}

\begin{figure}[p]
\epsfxsize19cm
\centerline{\epsffile{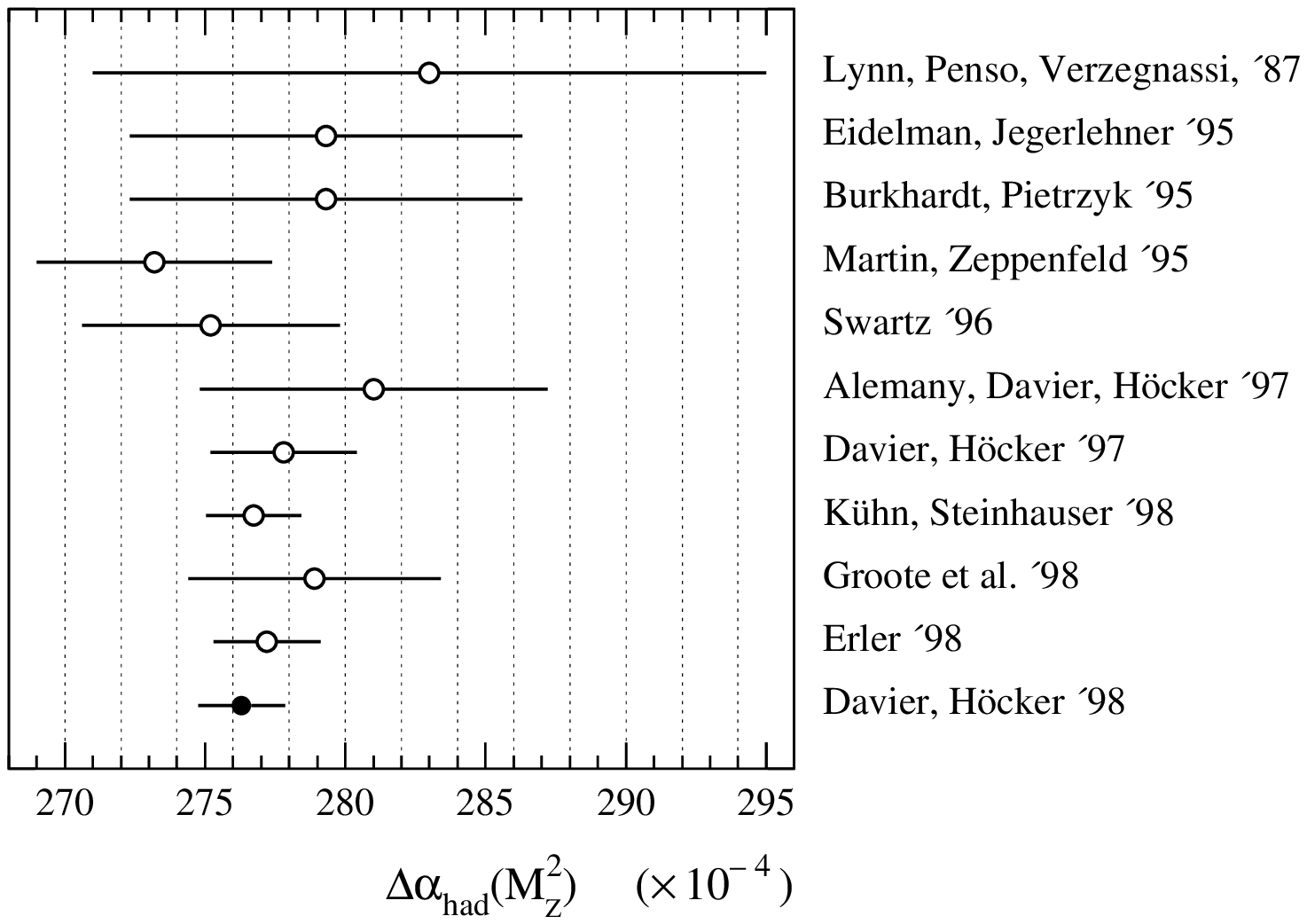}}
\caption[.]{\label{fig_results_alpha}\it 
            Comparison of \daqedhZ\ evaluations. The values are
            taken from Refs.~\rm\cite{lynn,eidelman,burkhardt,martin,swartz,
            g_2pap,alphapap,kuhnstein,erler} \it and from this work.}
\vspace{1.cm}
\epsfxsize19cm
\centerline{\epsffile{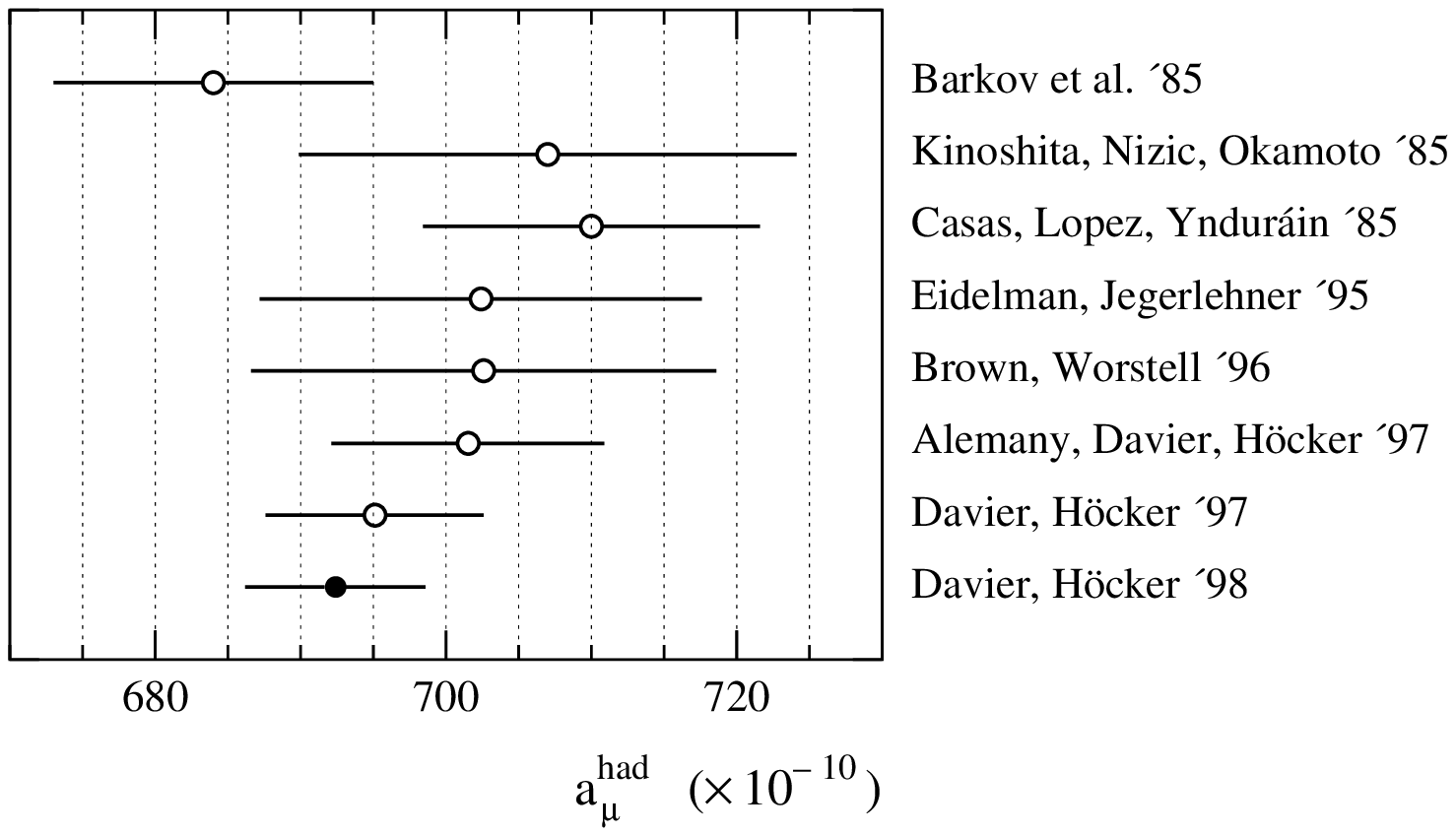}}
\caption[.]{\label{fig_results_amu}\it 
            Comparison of $a_\mu^{\rm had}$ evaluations. The values are
            taken from Refs.~\rm\cite{barkov,kinoshita,casas,eidelman,
            worstell,g_2pap,alphapap} \it and from this work.}
\end{figure}
We have reevaluated the hadronic vacuum polarization contribution to 
the running of the QED f\/ine structure constant, $\alpha(s)$, at
$s=M_{\rm Z}^2$ and to the anomalous magnetic moment of the muon,
$a_\mu$. We employed perturbative and nonperturbative 
QCD in the framework of the Operator Product Expansion in order
to extend the energy regime where theoretical predictions are 
reliable. Based on analyticity, we constrained the low-energy and 
$c\bar{c}$ threshold region theoretically using f\/inite energy 
sum rules and dispersion relations. The standard evaluation using 
data from \ee\ annihilation and $\tau$ decays at low energies and 
near quark production thresholds is therefore improved. Our results are
\daqedhZ\,$=(276.3\pm1.6)\times 10^{-4}$, propagating $\alpha^{-1}(0)$ 
to $\alpha^{-1}(M_{\rm Z}^2)=(128.933\pm0.021)$, and 
$a_\mu^{\rm had}=(692.4\pm6.2)\times10^{10}$ which yields the Standard 
Model prediction $a_\mu^{\rm SM}=(11\,659\,159.5\pm6.7)\times10^{-10}$.
For the electron we found $a_e^{\rm had}=(187.5\pm1.8)\times 10^{-14}$.
These results have direct implications for on-going experimental 
programs. On one hand, the precision on \aqedZ\ is now such that it
does not limit anymore the adjustment of the Higgs mass from 
accurate experimental determinations of 
${\rm sin}^2\theta_{\rm W}$~\cite{alphapap}. On the other hand, the
gain in accuracy for $a_\mu^{\rm had}$ is even more rewarding as it
will permit to exploit the foreseen precision increase of the $a_\mu$
measurement~\cite{bnl} in order to achieve a signif\/icant determination 
of the contribution $a_\mu^{\rm weak}$.

%
%


\begin{thebibliography}{99}
\bibitem{kuhnpriv}   We are indebted to J.~K\"uhn for helpful information
                     on this point
\end{thebibliography}
\end{document}